\documentclass[article,reqno]{amsart}

\usepackage{amssymb}
\usepackage{latexsym}
\usepackage{epsfig}
\usepackage{cite}
\usepackage{booktabs}
\usepackage{color}
\usepackage{amsthm}
\usepackage{graphicx}
\usepackage{tikz}
\usepackage{pgfplots}

\newcommand{\eqn}{\begin{eqnarray}}
\newcommand{\feqn}{\end{eqnarray}}
\newcommand{\beq}{\begin{equation}}
\newcommand{\eeq}{\end{equation}}
\newcommand{\bes}{\begin{equation*}}
\newcommand{\ees}{\end{equation*}}
\newcommand{\smat}{\left( \begin{smallmatrix}}
\newcommand{\smct}{\end{smallmatrix}\right)}

\newtheorem{theorem}{Theorem}

\newcommand{\ident}{I}

\numberwithin{equation}{section}

\begin{document}

\title[$\pmb{\Phi-\Psi}$ model for Electrodynamics in dielectric media]{$\pmb{\Phi-\Psi}$ model for Electrodynamics in dielectric media: exact quantisation in the Heisenberg representation}
\author{Francesco Belgiorno$^1$, Sergio L Cacciatori$^{2,3}$, Francesco Dalla Piazza$^{4}$, Michele Doronzo$^2$}

\address{$^1$ Dipartimento di Matematica, Politecnico di Milano, Piazza Leonardo 32, 20133 Milano, Italy, and INdAM-GNFM, Italy}
\address{$^2$ Department of Science and High Technology, Universit\`a dell'Insubria, Via Valleggio 11, IT-22100 Como, Italy}
\address{$^3$ INFN sezione di Milano, via Celoria 16, IT-20133 Milano, Italy}
\address{$^4$ Universit\`a ``La Sapienza'', Dipartimento di Matematica, Piazzale A. Moro 2, I-00185, Roma,   Italy}


\begin{abstract} 
We investigate the quantization in the Heisenberg representation of a model which represents a simplification of the Hopfield model for dielectric media, 
where the electromagnetic field is replaced by a scalar field $\phi$ and the role of the polarization field  is played by a further scalar field $\psi$. 
The model, which is quadratic in the fields, is still characterized by a nontrivial physical content, as the physical particles correspond to the polaritons 
of the standard Hopfield model of condensed matter physics. Causality is also taken into account and a discussion of the standard interaction representation 
is also considered. 
\end{abstract}

\maketitle

\section{Introduction}
In recent years, the investigations on possible revelations of the Hawking effect in analogues realised in dielectric media, \cite{philbin,PRL2010-H,PRL2011-Answer,belgiorno-prd,NJP2011,finazzi-carusotto-pra13,PRL2013,finazzi-carusotto-pra14,jacquet}, 
have raised the necessity of disposing of a model describing the quantum electromagnetic field interacting with a dispersive medium, reproducing the typical phenomenological dispersion relations. 

We recall that in the Hopfield model, a purely phenomenological quantisation of the electromagnetic field in the dielectric medium is replaced by a 
picture where the electromagnetic field interacts with a set of oscillators reproducing sources for dispersive properties 
of the electromagnetic field in matter \cite{hopfield,fano,kittel,davydov}. We stress that we don't take into account absorption in our paper, which is reasonable as far as 
phenomena one is interested in are not too near the absorption region. Including absorption would imply 
a much more tricky approach (cf. e.g. \cite{barnett,suttorp-epl,suttorp-jpa}), which is far beyond the scope of the present paper.\\
Our model satisfies the requirement to be fully relativistic covariant. This is in agreement with the necessity to proper simulate electrodynamics of 
moving media \cite{minkowski,post,Penfield-Haus,gordon,balasz,phammauquan,synge,Watson-Jauch}, where phenomenological 
electrodynamics is adopted. Also in our case, in order to 
get a complete analysis, one needs to change the inertial frame passing, for example, from the frame where the medium is at rest, to the frame where a given signal is at rest, or to the lab frame if it does not coincide with one of them.
With this in mind, a set of models taking into account the dispersion relations have been developed, based on a covariant reformulation of the Hopfield model \cite{PRD2015,Unp2015,EPJD2014,PhysicaScripta}.
In particular, in \cite{PRD2015} the Hopfield model has been presented in a simplified version, where, in  a two dimensional model, the electromagnetic field has been replaced by a massless scalar field, linearly coupled to a 
polarisation field, represented by a field of oscillators with characteristic pulsation $\omega_0$. This had the aim of simplifying several technical complications, but keeping the main relevant characteristics, as Lorentz covariance and the right
dispersion relation. In the latter reference, we were interested in the Hawking effect and we did not focus on a systematic quantisation of the model. 
A perturbative quantisation in a given gauge and in the lab frame of the whole homogeneous and 
isotropic relativistic Hopfield model has been presented in \cite{EPJD2014}, and a Lorentz and gauge covariant perturbative quantisation has been provided in \cite{PhysicaScripta}. However, this model can be quantised exactly, the exact
quantisation is involved with 
some significant physical characteristics of the given system, and in particular in its spectral properties, so quantizing non-perturbatively in the Heisenberg representation 
of quantum field theory is far more than a simple and straightforward exercise. The exact quantisation of the relativistic Hopfield model will be presented elsewhere
\cite{Quantum}: its construction is quite involved and passes through several technical intricacies, that go beyond the ones characterising the present model.\\
Since the main steps characteristic of the model are present also in the scalar simplified analogue, without the intricacies due to gauge invariance and the presence of unphysical modes, we will present here the exact quantisation
of the scalar Hopfield model, which we dub the \pmb{$\Phi-\Psi$} model. Since our aim is to illustrate here the strategy of \cite{Quantum} without hiding it behind technical details, we will mainly present the results without proves, illustrating the 
main steps at an intuitive level, by mean of precise statements, which will be fully proved, in a more general form, in \cite{Quantum} and \cite{Path-Hopfield}.

\

In section \ref{sec:quantisation} we present the model and its quantisation, by showing how the Fock representation for the whole interacting model can be realised. In section \ref{sec:causality and covariance} we show how the problem of
causality is related to Lorentz covariance and show how it can be realised in our model, which is only covariant and not invariant under Lorentz transformations. In section \ref{sec: propagator} we compute the propagator both directly from the 
Fock representation and with the path integral method, and state their equivalence. In section \ref{sec:fano} we discuss the 
interaction representation and the Fano diagonalization method. 
In section \ref{sec: final} we add some further discussion. All statements are simple consequences of the ones proved in \cite{Quantum} and \cite{Path-Hopfield}.

\section{The $\pmb{\Phi-\Psi}$ model and its quantisation}\label{sec:quantisation}
We consider the $D+1$ dimensional $\pmb{\Phi-\Psi}$ model whose classical dynamics is defined by the action
\begin{eqnarray}
S[\phi,\psi] &=& \int d^{D+1}x \left[ \frac 12 \partial_\mu \phi \partial^\mu \phi +\frac 12 v^\mu \partial_\mu \psi v^\nu \partial_\nu \psi -\frac {\omega_0^2}2 \psi^2-g \phi v^\mu \partial_\mu \psi \right] =\int d^{D+1} x\ \mathcal L_c,
\end{eqnarray}
where $\pmb v$ is the spacetime velocity of the rest frame for the $\psi$ field. The conjugate momenta are
\begin{eqnarray}
&&\pi_\phi=\partial_t \phi, \label{pifi}\\
&&\pi_\psi=v^0v^\mu \partial_\mu \psi- gv^0\phi,\label{pipsi}
\end{eqnarray}
so that the Hamiltonian is
\begin{eqnarray}
H=\int_{\mathbb R^{D}}\ d^D\vec x &&\left[\frac{\pi_\phi^2}{2} +\frac{\pi_\psi^2}{2v_0^2} +\frac{\pi_\psi}{v_0} \left( g\phi- \frac {\vec v}{v^0}\cdot \vec \nabla \psi \right) +\frac 12 \vec \nabla \phi \cdot \vec \nabla \phi +\frac {g^2}2 \phi^2
+\frac {\omega_0^2}2 \psi^2 \right].
\end{eqnarray}
The classical equations of motion in the Fourier space are
\begin{eqnarray}
\pmb {\mathcal M} V\equiv 
\left(
\begin{array}{cc}
-k^2 & -ig\omega \\
ig\omega & -\omega^2+\omega_0^2  
\end{array}
\right)
\left(
\begin{array}{c}
\tilde \phi \\ 
\tilde \psi 
\end{array}
\right)
=
\left(
\begin{array}{c}
0 \\  
0
\end{array}
\right),
\end{eqnarray} \label{equation motion}
where $\omega:=k^\mu v_\mu$. The dispersion relation is given by  $\det \mathcal M=0$, that is
\begin{eqnarray}
DR(\pmb k):=k^2 -\frac {g^2 \omega^2}{\omega^2-\omega_0^2}=0.
\end{eqnarray}
This defines the support of the solutions in the momentum space, with two positive branches, corresponding to the two solutions having positive $\omega$, which we will indicate with $k^0_{(a)}$, $a=1,2$. Everywhere the suffix $(a)$ will
mean ``evaluated at $k^0=k^0_{(a)}(\vec k)$''. We will also use the symbol $\vec k$ for the spatial component of a spacetime vector $\pmb k$, and similar for all vectors, whereas $k^2:= \pmb k\cdot \pmb k$.\\
The classical solutions of the equations of motion are
\begin{eqnarray}
&&\phi(\pmb x)=\sum_{a=1}^2 \int_{\mathbb R^D} d\mu (\vec{k}) \left[ a_{(a)}(\vec k) e^{-i\pmb k_{(a)}\cdot \pmb x} +a^\dagger_{(a)}(\vec k) e^{i\pmb k_{(a)}\cdot \pmb x}  \right], \\
&&\pi_\phi (\pmb x)=\sum_{a=1}^2 \int_{\mathbb R^D} d\mu (\vec{k}) (-i k^0_{(a)})  \left[ a_{(a)}(\vec k) e^{-i\pmb k_{(a)}\cdot \pmb x} -a^\dagger_{(a)}(\vec k) e^{i\pmb k_{(a)}\cdot \pmb x}  \right],\\
&&\psi(\pmb x)=\sum_{a=1}^2 \int_{\mathbb R^D} d\mu (\vec{k}) \frac {ig\omega_{(a)}}{\omega_{(a)}^2-\omega_0^2} \left[ a_{(a)}(\vec k) e^{-i\pmb k_{(a)}\cdot \pmb x} -a^\dagger_{(a)}(\vec k) e^{i\pmb k_{(a)}\cdot \pmb x}  \right], \\
&&\pi_\psi (\pmb x)=\sum_{a=1}^2 \int_{\mathbb R^D} d\mu (\vec{k}) \frac {gv^0 \omega_0^2}{\omega_{(a)}^2-\omega_0^2}  \left[ a_{(a)}(\vec k) e^{-i\pmb k_{(a)}\cdot \pmb x} +a^\dagger_{(a)}(\vec k) e^{i\pmb k_{(a)}\cdot \pmb x}  \right],
\end{eqnarray}
where
\begin{eqnarray}
d\mu (\vec{k}):= \frac {d^D\vec k}{(2\pi)^D} \frac 1{DR'_{(a)}},
\end{eqnarray}
and 
\begin{eqnarray}
DR'_{(a)}(\vec k):= \frac {dDR}{dk^0} (\pmb k_{(a)})=2k^0_{(a)} +2\frac {g^2\omega_0^2\omega_{(a)} v^0}{(\omega_{(a)}^2-\omega_0^2)^2}.
\end{eqnarray}
The set of such functions 
\begin{eqnarray}
\Psi=
\left(
\begin{array}{c}
\phi \\ \psi \\ \pi_\phi \\ \pi_\psi
\end{array}
\right)
\end{eqnarray}
is endowed with the conserved scalar product
\begin{eqnarray}
(\Psi_1|\Psi_2)=i \int_{\mathbb R^D} d^D\vec x \Psi_1^* (\pmb x) \Omega \Psi_2 (\pmb x), 
\end{eqnarray}
with
\begin{eqnarray}
\Omega=
\left( \begin{array}{cc}
\mathbb O_2 & \mathbb I_2 \\
-\mathbb I_2 & \mathbb O_2
\end{array} 
\right).
\end{eqnarray}
A basis of positive norm plane waves is 
\begin{eqnarray}
\zeta_{(a)}(\vec k;\pmb x)=e^{-i\pmb k_{(a)}\cdot \pmb x} 
\left( \begin{array}{c}
1 \\ ig \frac {\omega_{(a)}}{\Omega_{(a)}^2-\omega_0^2} \\ -ik^0_{(a)} \\ gv^0 \frac {\omega_0^2}{\omega_{(a)}^2-\omega_0^2}
\end{array} 
\right).
\end{eqnarray}
Notice that
\begin{eqnarray}
(\zeta_{(a)}(\vec k)| \zeta_{(b)}(\vec q))=\delta_{ab} \delta^D(\vec k-\vec q) (2\pi)^D DR'_{(a)}(\vec k),
\end{eqnarray}
which gives
\begin{eqnarray}
a_{(a)}(\vec k)=(\zeta_{(a)}(\vec k)|\Psi)
\end{eqnarray}
and then
\begin{eqnarray}
[a_{(a)}(\vec k),a_{(b)}^\dagger(\vec q)]=(\zeta_{(a)}(\vec k)|\zeta_{(b)}(\vec q))=\delta_{ab} \delta^D(\vec k-\vec q) (2\pi)^D DR'_{(a)}(\vec k),
\end{eqnarray}
and all other commutators vanish. This result has been obtained by imposing that the quantum fields satisfy the equal time canonical commutation relations, which, as a consequence, are satisfied. Nevertheless, it is interesting to point out that the
validity of the last ones is consequence of a series of nontrivial identities that we summarise in an appendix. This way, one can proceed in the usual way in constructing the Fock space, starting from the vacuum state $\Omega$, the unique normalised
state that is annihilated by all $a_{(a)}(\vec k)$, we can realise the Fock space as the completion of the set of states  generated by all polynomial actions of the creator fields $a^\dagger_{(a)}(\vec k)$. This is standard and free of 
particular difficulties, apart from the fact that the vacuum state is unique only after fixing a choice of $\pmb v$, since the theory is not invariant under the whole Poincar\'e group, but only under the subgroup leaving $\pmb v$ invariant.

Here, we simply notice that the fact that the CCR are satisfied, together with Lorentz covariance, allow to prove that the principle of causality is satisfied. Since the explicit presence of $\pmb v$ breaks the Lorentz invariance, the question
of the covariance is a little bit delicate and requires a careful analysis. 

\section{Causality and covariance}\label{sec:causality and covariance}
The algebra of quantum fields is generated by the canonical commutation relations (CCR):
\begin{eqnarray}
&&[\phi(t,\vec x),\pi_\phi(t,\vec y)]=i\delta^D(\vec x-\vec y), \qquad [\psi(t,\vec x),\pi_\psi(t,\vec y)]=i\delta^D(\vec x-\vec y), \nonumber
\end{eqnarray}
where we indicated only the non-zero contributions. 
Causality conditions are apparently a little bit stronger:
\begin{eqnarray}
&&[\phi(\pmb x),\phi(\pmb y)]= [\phi(\pmb x),\psi(\pmb y)]=[\phi(\pmb x),\pi_\psi(\pmb y)]=[\psi(\pmb x),\psi(\pmb y)]= 0, \cr
&&[\psi(\pmb x),\pi_\phi(\pmb y)]= [\pi_\phi(\pmb x),\pi_\phi(\pmb y)]=[\pi_\phi(\pmb x),\pi_\psi(\pmb y)]= [\pi_\psi(\pmb x),\pi_\psi(\pmb y)]=0, \label{causal} \\
&&[\phi(\pmb x),\pi_\phi(\pmb y)]=[\psi(\pmb x),\pi_\psi(\pmb y)]=0, \nonumber
\end{eqnarray}
for any pair of points $\pmb x, \pmb y$ spatially separated, $(\pmb x-\pmb y)^2<0$. If the theory is Lorentz invariant, then (\ref{causal}) follow from the CCRs, since we can change frame into the one where $x^0=y^0=t$
and then employ the CCRs in order to prove their vanishing. 
Our model Lagrangian is associated with equations of motion which are covariant with respect to the Lorentz group. This fact does not correspond to a full Lorentz invariance, due to the fact that the Lorentzian metric is not the only absolute object of the theory, 
but a further absolute object \cite{anderson} appears: the velocity $\pmb v$ of the medium. This implies that our theory, and any covariant theory of a dielectric medium (cf. e.g. \cite{jauch1}), is involved with a preferred frame, which corresponds to the rest frame 
of the medium.
The explicit presence  of the vector $\pmb v$ in the Lagrangian implies that, in general, boosts are no more symmetries. This breaks the Poincar\'e symmetry group down to the subgroup leaving $\pmb v$ invariant. This behaviour should not 
be a surprise, as it is common to all the cases where e.g. Klein-Gordon equation is studied in presence of an external potential. 
Loss of Lorentz invariance is evident, but, at the same time, the field equations are covariant, and solutions are transformed into solutions of the Klein-Gordon equation by the Lorentz group, provided that external potential is transformed 
too. See e.g. \cite{gerard} (p. 516).  In this sense, also our theory remains Lorentz covariant, as the equations of motions are, and the above argument proving causality applies again if the covariance is respected at the level of the representation 
of the quantum theory. 
Covariance ensures that different inertial observers perceive the same physics, i.e. are involved with the same processes with the same 
probability. In particular, the number of polaritons in the process remain the same. Unitary maps between Fock spaces of different inertial observer 
are a natural consequence of Poincar\'e covariance, and is a consequence of the fact that the Lagrangian is invariant under simultaneous transformations 
of the fields and of the vector $\pmb v$ under Poincar\'e group. So, at the quantum level
there exists a Fock space $\mathcal F_{\pmb v}$ for any $\pmb v$ and a set of unitary maps
\begin{eqnarray}
U: &&G\longrightarrow \mathcal {U}, \\
&& \Lambda  \longmapsto U(\Lambda) : \mathcal {F}_{\pmb v} \longrightarrow \mathcal {F}_{U(\Lambda) \pmb v}, \ \forall \pmb v\ {\mathrm{ timelike}},
\end{eqnarray}
where $\mathcal {U}$ is the set of all possible isometric maps among Fock spaces $\mathcal {F}_{\pmb v}$ and $\mathcal {F}_{\pmb w}$, and $U(\Lambda)$ is 
defined by
\begin{eqnarray}
U_{\pmb v}(\Lambda)|0\rangle_{\pmb v}=|0\rangle_{\Lambda \pmb v}, 
\end{eqnarray}
and
\begin{eqnarray}
U_{\pmb v}(\Lambda) o(\vec k; \pmb v) U_{\pmb v}(\Lambda)^{-1}=o(U(\Lambda)\vec k; U(\Lambda)\pmb v),
\end{eqnarray}
where $o$ is intended to be any one among the operators $a_{(a)}(\vec k), a^\dagger_{(a)}(\vec k)$, $a=1,2$. 


\section{The propagator}\label{sec: propagator}
Since the theory is Gaussian, it is completely determined by the two point functions. It is given by the matrix 
\begin{eqnarray}
iG^{IJ}_{\pmb v}(\pmb x,\pmb y)= {}_{\pmb v}\langle 0| T(\Phi^I(\pmb x)\Phi^J(\pmb y))|0\rangle_{\pmb v}, \qquad I,J=1,2,
\end{eqnarray}
where 
\begin{eqnarray}
\Phi^1=\phi, \quad \Phi^2=\psi.
\end{eqnarray}
We will also write
\begin{eqnarray}
G_{\pmb v}^{IJ}(\pmb x,\pmb y)=G_{\pmb v}^{IJ}(\pmb x,\pmb y)_+\theta (x^0-y^0)+G_{\pmb v}^{IJ}(\pmb x,\pmb y)_-\theta(y^0-x^0).
\end{eqnarray}
Since $G_{\pmb v}^{IJ}(\pmb x,\pmb y)_-$ is easily obtained from $G_{\pmb v}^{IJ}(\pmb x,\pmb y)_+$, we will write down only the latter: 
\begin{eqnarray}
\!\!\!\!\!\! iG_{\pmb v}^{11}(\pmb x,\pmb y)_+&=&{}_{\pmb v}\langle 0| \phi(\pmb x)\phi(\pmb y)|0\rangle_{\pmb v}=\sum_{a=1}^2 \int_{\mathbb R^D} d\mu (\vec{k}) 
{e^{-i\pmb k_{(a)}\cdot (\pmb x-\pmb y)} }; \label{fifi} \\
\!\!\!\!\!\! iG_{\pmb v}^{12}(\pmb x,\pmb y)_+&=&{}_{\pmb v}\langle 0| \phi(\pmb x)\psi(\pmb y)|0\rangle_{\pmb v}  = \sum_{a=1}^2\int_{\mathbb R^D} d\mu (\vec{k})\frac {-ig\omega_{(a)}}{\omega_{(a)}^2-\omega_0^2} 
{e^{-i\pmb k_{(a)}\cdot (\pmb x-\pmb y)} }; \label{fipsi}\\
\!\!\!\!\!\! iG_{\pmb v}^{22}(\pmb x,\pmb y)_+&=&{}_{\pmb v}\langle 0| \psi(\pmb x)\psi(\pmb y)|0\rangle_{\pmb v}
=\sum_{a=1}^2 \int_{\mathbb R^D} d\mu (\vec{k}) \frac {{g^2} \omega_{(a)}^2}{(\omega_{(a)}^2-\omega_0^2)^2} 
{e^{-i\pmb k_{(a)}\cdot (\pmb x-\pmb y)}}. \label{psipsi}
\end{eqnarray}
The propagator can be determined also by means of the path integral formulation.
After introducing the currents $J_\phi$ and $J_\psi$, we can define the functional generating the propagators:
\begin{eqnarray}
Z[J_\phi, J_\psi]=\pmb \int [D\phi D\psi]  \exp &&\left\{i\int_{\mathbb R^{D+1}} \mathcal{L}_c d^{D+1}\pmb x + i \int_{\mathbb R^{D+1}} J_{\phi} \phi d^{D+1} \pmb x + i \int_{\mathbb R^{D+1}} J_{\psi} \psi d^{D+1} \pmb x
\right\}.
\end{eqnarray}
From this we can formally compute the propagator, which results to be
\begin{eqnarray}
\!\!\!\!\!\! G_{\Phi^I \Phi^J}(\pmb x,\pmb y)=\left.-i \frac {\delta Z[J_\phi, J_\psi]}{\delta J_{\Phi^I}(\pmb x) \delta J_{\Phi^J}(\pmb y)}\right|_{\pmb J_{K}}= \int_{\mathbb R^2} \frac {d^{D+1} \pmb k}{(2\pi)^{D+1}} e^{-i\pmb k\cdot (\pmb x-\pmb y)} \pmb {\mathcal M}^{-1} 
(\pmb k)_{IJ},
\end{eqnarray}
where $\pmb {\mathcal M}$ is defined in (\ref{equation motion}), so that 
\begin{eqnarray}
\pmb {\mathcal M}^{-1} (\pmb k)=\frac 1{k^2(\omega^2-\omega_0^2)-g^2\omega^2}
\left( \begin{array}{cc}
-\omega^2+\omega_0^2  & ig\omega \\
-ig\omega & -k^2
\end{array} 
\right).\label{M-1}
\end{eqnarray} 

Naturally, this is not the complete story, since this expression requires a prescription avoiding the poles defined by the dispersion relation. Such a prescription 
must respect causality. It results that this can be accomplished by means of a $i\varepsilon$ Feynman prescription. Indeed, it holds:
\begin{theorem}\label{proposition}
The propagator is
\begin{eqnarray}
G_{\Phi^I \Phi^J}(\pmb x, \pmb y)= \int_{\mathbb R^{D+1}} \frac {d^{D+1}\pmb k}{(2\pi)^{D+1}}e^{-i\pmb k \cdot (\pmb x-\pmb y)} \pmb {\mathcal M}^{-1}_{i\varepsilon} (\pmb k)_{IJ}, \label{propagatore}
\end{eqnarray}
where $\pmb{\mathcal M}^{-1}_{i\varepsilon} (\pmb k)$ is obtained from (\ref{M-1}) by taking the complex shifts $k^2 \to k^2+i\varepsilon$, $\omega_0^2\to \omega_0^2-ic^2\varepsilon$.
\end{theorem}
This proposition is a particular case of a more general one proved in \cite{Path-Hopfield}.

\section{The interaction representation and Fano diagonalization}\label{sec:fano}

In this section we take into account the more standard Interaction representation, and perform the so-called Fano diagonalization \cite{fano} of the full Hamiltonian 
operator in order to find its eigenmodes. As the Hamiltonian is quadratic in the fields and their conjugate momenta, we are able to obtain an exact 
result which leads again to polaritons as physical states of the system. It is interesting to stress that this approach, which is pursued both in the 
original paper by Hopfield \cite{hopfield} and in standard textbooks (see e.g. \cite{kittel,davydov}), is in principle apt to perturbation theory and 
leads to the same result thanks to the diagonalization process.\\
In line of principle, the Interaction representation is constructed by assuming that $g$ is small and allows a perturbation theory in powers of 
$g$. For simplicity, we consider only the case where $\pmb{v}=(c,\vec{0})$, and we put $c=1$.  
The Hamiltonian is then characterized by three contributions: two free-fields contributions $H^0_\phi,H^0_\psi$, 
and an interaction one $H_{int}$ as in the following equations
\begin{eqnarray}
H^0_\phi &=&
\int_{\mathbb R^{D}}\ d^D\vec x \left[\frac{\pi_\phi^2}{2} +\frac 12 \vec \nabla \phi \cdot \vec \nabla \phi\right],\\
H^0_\psi &=& \int_{\mathbb R^{D}}\ d^D\vec x \left[\frac{\pi_\psi^2}{2v_0^2} +\frac {\omega_0^2}2 \psi^2 \right],\\
H_{int} &=&\int_{\mathbb R^{D}}\ d^D\vec x \left[g\phi \frac{1}{v^0}\pi_\psi+\frac {g^2}2 \phi^2\right].
\end{eqnarray}
Note that the conjugate momentum $\pi_\psi$ is now a free field one (i.e. is the one in (\ref{pipsi}) with $g=0$), and that the dispersion 
relation for the free field $\phi$ is $(k^0)^2-\vec{k}\cdot \vec{k}=0$, whereas the dispersion relation for $\psi$ is $\omega^2-\omega_0^2=0$. 
We now write the free fields in terms of creation and annihilation operators. In order to allow a more direct comparison with the existing literature, we 
choose 
\begin{eqnarray}
&&\phi(\pmb x)=\int_{\mathbb R^D} \frac{d^D\vec k}{(2\pi)^{D/2}} \frac{1}{\sqrt{2 k^0}} \left[ b(\vec k,t) e^{-i\vec k \cdot \vec x} +h.c.\right],\\
&&\pi_\phi (\pmb x)=\int_{\mathbb R^D} \frac{d^D\vec k}{(2\pi)^{D/2}} \sqrt{\frac{k^0}{2}} \left[-i b (\vec k,t) e^{-i\vec k \cdot \vec x}  +h.c.\right],\\
&&\psi(\pmb x)= \int_{\mathbb R^D} \frac{d^D\vec k}{(2\pi)^{D/2}} \frac{1}{\sqrt{2\omega_0}} \left[ d(\vec k,t) e^{-i\vec k \cdot \vec x} +h.c.\right],\\
&&\pi_\psi (\pmb x)=\int_{\mathbb R^D} \frac{d^D\vec k}{(2\pi)^{D/2}} \sqrt{\frac{\omega_0}{2}}  \left[ -i d(\vec k,t) e^{-i\vec k \cdot \vec x} +h.c.\right],
\end{eqnarray}
with
\begin{eqnarray}
&&[b (\vec{k},t),b^\dagger (\vec{q},t)] =\delta^D (\vec{k}-\vec{q}),\\
&&[d (\vec{k},t),d^\dagger (\vec{q},t)] =\delta^D (\vec{k}-\vec{q}),
\end{eqnarray}
and all the remaining CCRs equal to zero. We obtain 
\begin{eqnarray}
H^0_\phi   &=&
\int d^D\vec p\ p_0\ b^\dagger (\vec{p},t) b (\vec{p},t),\\
H^0_\psi &=& \int d^D\vec p\ \omega_0\ d^\dagger (\vec{p},t) d (\vec{p},t),\\
H_{int} &=& \int d^D\vec p\ \left[ -i \frac{g}{2} \sqrt{\frac{\omega_0}{p_0}} \left(d (\vec{p},t) b^\dagger (\vec{p},t)- d^\dagger  (\vec{p},t) b (\vec{p},t)+d (\vec{p},t) b (-\vec{p},t)-d^\dagger  (\vec{p},t) b^\dagger (-\vec{p},t)
\right) \right.\cr
&& \left.  \hphantom{ \int d^D\vec p\ -i  }+ \frac{g^2}{4p_0} \left(
b (\vec{p},t) b^\dagger (\vec{p},t)+b^\dagger  (\vec{p},t) b (\vec{p},t) +b (\vec{p},t) b (-\vec{p},t)+b^\dagger  (\vec{p},t) b^\dagger (-\vec{p},t)
\right) \right].
\end{eqnarray}
The diagonalization process consists in finding normal modes annihilation operators 
\begin{equation}
\alpha (\vec{p},t) = w\ b(\vec{p},t)+ x\ d(\vec{p},t) + y\ b^\dagger (-\vec{p},t)+ z\ d^\dagger (-\vec{p},t)
\end{equation}
such that \cite{hopfield}
\begin{equation}
\left[\alpha (\vec{p},t), H\right] = E(\vec{p}) \alpha (\vec{p},t).
\end{equation}
The former eigenvalue problem amounts to the following one:
\begin{equation}
 \det [ A - E \ident ]=0, \label{eigen}
\end{equation}
where $A$ is the matrix
\begin{equation}
A =\left[
\begin{array}{cccc}
p_0+ \frac{g^2}{2 p_0} & i \frac{g}{2} \sqrt{\frac{\omega_0}{p_0}} & - \frac{g^2}{2 p_0} & i \frac{g}{2} \sqrt{\frac{\omega_0}{p_0}} \cr
-i \frac{g}{2} \sqrt{\frac{\omega_0}{p_0}} & \omega_0 & i \frac{g}{2} \sqrt{\frac{\omega_0}{p_0}} & 0\cr
 \frac{g^2}{2 p_0} & i \frac{g}{2} \sqrt{\frac{\omega_0}{p_0}} & -p_0 - \frac{g^2}{2 p_0} &i \frac{g}{2} \sqrt{\frac{\omega_0}{p_0}} \cr
i \frac{g}{2} \sqrt{\frac{\omega_0}{p_0}} & 0 &- i \frac{g}{2} \sqrt{\frac{\omega_0}{p_0}} &-\omega_0
\end{array}
\right]
\end{equation}
From (\ref{eigen}) one obtains the equation
\begin{equation}
E^4 - E^2 (|\vec{k}|^2 +g^2 +\omega_0^2) +|\vec{k}|^2 \omega_0^2=0,
\end{equation}
which amounts to 
\begin{equation}
|\vec{k}|^2 = E^2 \left( 1+ \frac{g^2}{\omega_0^2 -E^2} \right),
\end{equation}
which again gives the same eigenmodes as in the previous sections. In particular, two positive branches $E_\pm$ can be obtained, 
with associated eigenvectors. One can obtain $\alpha_a$, with $a=\pm$, which correspond to the  $a_{(a)}$ given in the previous analysis. 

\section{Final comments}\label{sec: final}
In the appendix it is shown that $DR'_{(a)}$, $a=1,2$, are always positive, apart in $\vec k=0$ for the lower branch. In this case, $DR'$ vanishes linearly in $|\vec k|$, and the integrals defining the fields (in a distributional sense) are well posed for $D>1$.
Thus, the fields and propagators are well defined as tempered distributions.\\
On the opposite, in a two dimensional spacetime, the integrals diverge unless the oscillator modes vanish quickly enough at the origin. This divergence at $k=0$ cannot be interpreted as an infrared divergence, since it does not occur only
in the propagator, but also in the definition of the fields, so that it needs to be eliminated by a suitable choice of the space of test functions. Assuming that these must be chosen inside the set of smooth rapidly decreasing functions $S(\mathbb R^2)$,
we must consider functions whose Fourier transform vanishes in $\vec k=0$ when $k^0$ is evaluated on $k^0_{(2)}(k)$. Since also $k^0$ vanishes, by employing smoothness, it is sufficient to consider functions in $S(\mathbb R^2)$ whose
Fourier transform vanish in $\pmb k=\pmb 0$. These are the rapidly decreasing smooth functions having null mean. So, the test functions must be chosen in
\begin{eqnarray}
S_0(\mathbb R^2)=\left\{f\in S(\mathbb R^2)| \int_{\mathbb R^2} d^2 \pmb x f(\pmb x)=0 \right\}.
\end{eqnarray}
Thus, it is exactly the same as for any massless free field.\\ 
After having defined the fields, we have been able to treat them exactly, being the action quadratic.
However, despite the theory is Gaussian and, then, essentially a free theory, in a sense the interaction manifests itself through a highly nontrivial dispersion relation.\\ 
The second difficulty, related to causality, is covariance, which is realised in a nontrivial way. We have just sketched how this can be done: in practice the quantum algebra is represented on a bundle of Fock spaces over the homogeneous
space\footnote{Since $\pmb v$ is assumed to be a future directed timelike vector of norm 1, and $P$ acts transitively on the set of such vectors, then 
$$\mathcal B\simeq \{ \pmb x\in \mathbb R^{1,3}| x^2=1, x^0>0 \},$$ that is the future paraboloid of mass 1 in the Minkowski space $\mathbb R^{1,3}$.}
\begin{eqnarray}
\mathcal B=\mathcal P/G,
\end{eqnarray}
where $\mathcal P$ is the Poincar\'e group and $G$ its subgroup that leaves $\pmb v$ invariants (the little group of $\pmb v$). \\
Finally, we have computed the two point function, which characterises the whole theory, being it Gaussian. We have done it both starting from the canonical representation and with the path integral method. Here, the $i\varepsilon$
Feynman-St\"uckelberg prescription has been introduced and stated to be equivalent to the causal propagator computed in the oscillator representation. The proof can be found in \cite{Path-Hopfield}. \\
In the relativistic Hopfield model, the true target of all our efforts, all this difficulties are present and amplified by the presence of an higher number of field components, including non physical ones, a larger number of spectral branches,
the necessity of taking under control the gauge symmetry, the presence of constraints and of the dipole ghost, and a major involution of all explicit formulas. However, all this additive complications, which require to be overcome, are not
 peculiar of the specific model and ends up in hiding the specific ones, which are transparent in the simplified $\pmb {\Psi-\Phi}$ model we have presented here.

\newpage
\begin{appendix}\label{app: relations}

\section{On the dispersion relation}
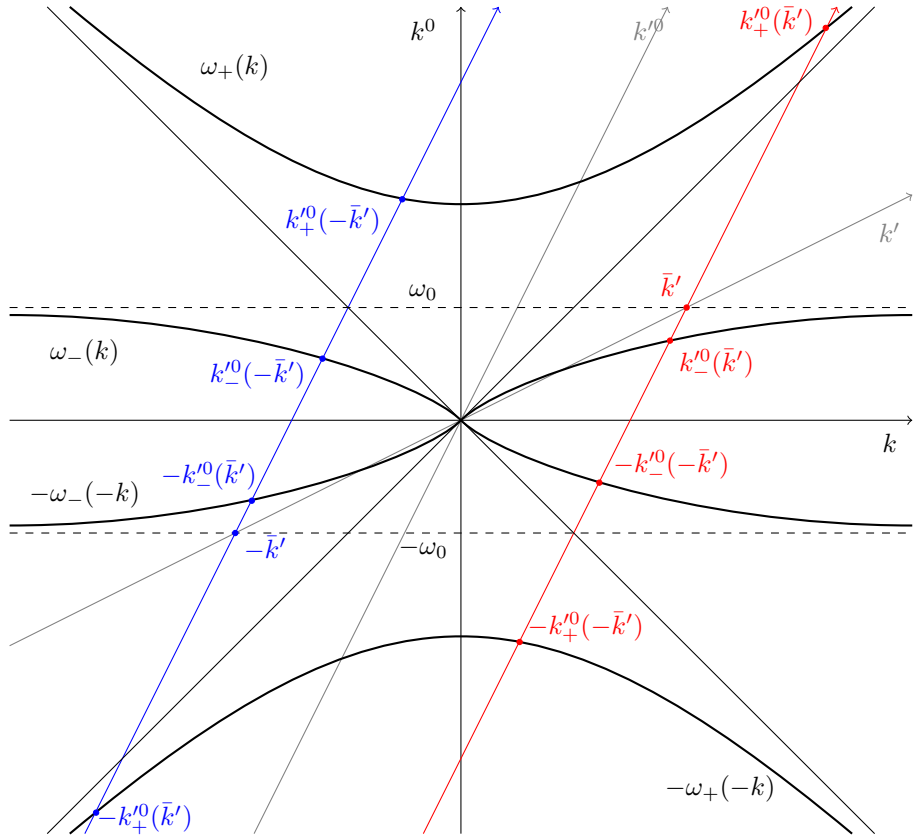
\begin{figure}[!htbp]
\begin{center}
\begin{tikzpicture} 
\draw [->](-1,0) -- (11,0);
\draw [->](5,-5.5) -- (5,5.5);
\draw[gray] [->](-1,-3) -- (11,3);
\draw[gray] [->](2.25,-5.5) -- (7.75,5.5);
\draw[red] [->](4.5,-5.5) -- (10,5.5);
\draw[blue] [->](0,-5.5) -- (5.5,5.5);
\draw [-,dashed](-1,1.5) -- (11,1.5);
\draw [-,dashed](-1,-1.5) -- (11,-1.5);
\draw [-](-0.5,-5.5) -- (10.5,5.5);	
\draw [-](-0.5,5.5) -- (10.5,-5.5);
\draw[thick] (-0.2,5.5) .. controls (4,2) and (6,2) .. (10.2,5.5);
\draw[thick] (-0.2,-5.5) .. controls (4,-2) and (6,-2) .. (10.2,-5.5);
\draw[thick] (5,0) .. controls (5.5,0.5) and (8,1.4) .. (11,1.4);
\draw[thick] (5,0) .. controls (4.5,-0.5) and (2,-1.4) .. (-1,-1.4);
\draw[thick] (5,0) .. controls (5.5,-0.5) and (8,-1.4) .. (11,-1.4);
\draw[thick] (5,0) .. controls (4.5,0.5) and (2,1.4) .. (-1,1.4);
\filldraw [red]
(8,1.5) circle (1pt)
(7.78,1.06) circle (1pt)
(9.85,5.22) circle (1pt)
(6.84,-0.83) circle (1pt)
(5.78,-2.95) circle (1pt);
\filldraw [blue]
(2,-1.5) circle (1pt)
(2.22,-1.07) circle (1pt)
(0.15,-5.22) circle (1pt)
(3.16,0.82) circle (1pt)
(4.22,2.94) circle (1pt);
\node[red] at (7.8,1.8) {$\bar k'$};
\node[red] at (9.2,5.3) {$k_+^{\prime 0}(\bar k')$};
\node[red] at (8.4,0.75) {$k_-^{\prime 0}(\bar k')$};
\node[red] at (7.8,-0.6) {$-k_-^{\prime 0}(-\bar k')$};
\node[red] at (6.66,-2.75) {$-k_+^{\prime 0}(-\bar k')$};
\node[blue] at (2.4,-1.7) {$-\bar k'$};
\node[blue] at (0.8,-5.3) {$-k_+^{\prime 0}(\bar k')$};
\node[blue] at (1.65,-0.75) {$-k_-^{\prime 0}(\bar k')$};
\node[blue] at (2.3,0.6) {$k_-^{\prime 0}(-\bar k')$};
\node[blue] at (3.3,2.65) {$k_+^{\prime 0}(-\bar k')$};
\node at (4.5,1.7) {$\omega_0$};
\node at (4.5,-1.7) {$-\omega_0$};
\node at (4.5,5.2) {$k^0$};
\node at (10.7,-0.3) {$k$};
\node at (2,4.7){$\omega_+(k)$};
\node at (0,0.9){$\omega_-(k)$};
\node at (8.45,-4.9){$-\omega_+(-k)$};
\node at (0,-0.99){$-\omega_-(-k)$};
\node[gray] at (7.5,5.2) {$k^{\prime0}$};
\node[gray] at (10.7,2.5) {$k'$};
\end{tikzpicture}
\caption{The thick black lines are the solutions of dispersion relations in the lab frame. Above the upper one $DR(\pmb k)$ is positive. Below it is negative until reaching the asymptote $k^0=\omega_0$. Below it $DR$ is again positive
and it changes sign in crossing the line $k^0=\omega_-(k)$. This behaviour continues symmetrically in the lower half plane. The grey lines represent the axes of a boosted frame and the red line is at fixed $k'$. It cuts the curves $DR=0$
from increasing values of $DR$ (when $k^{\prime0}$ increases) in the upper half plane and with decreasing sign in the lower half plane.}\label{figura dispersion}
\end{center}
\end{figure}

It may be useful to analyse some properties of the dispersion relations. They can be exactly solved in the lab frame (defined by $\pmb v\equiv(1,\vec 0)$). There are four real solutions for $k^0$, which are 
$k_{(1)}^0=\omega_+(\vec k)$, $k_{(2)}^0=\omega_-(\vec k)$, $k_{(3)}^0=-\omega_-(-\vec k)$, $k_{(4)}^0=-\omega_+(-\vec k)$, with
\begin{eqnarray}
\omega_\pm(\vec k)=\frac 12 \sqrt{(\omega_0+|\vec k|)^2+g^2} \pm\frac 12 \sqrt{(\omega_0-|\vec k|)^2+g^2}.
\end{eqnarray}
These four branches, two positive and two negative, are represented in figure \ref{figura dispersion} (in the $D=1$ case), from which it is also evident that $DR'_{(a)}$ is positive for $a=1,2$ and negative for $a=3,4$, and vanishes only in the origin,
thus for $\vec k=0$ in the branches $2,3$. In this limit we see that 
\begin{eqnarray}
\omega_-=\frac {\omega_0}{\omega_+} |\vec k|\approx \frac {\omega_0}{\sqrt {\omega_0^2+g^2}} |\vec k|,
\end{eqnarray}
and, then, $DR'$ vanishes linearly in $|\vec k|$.

\end{appendix}

\newpage

\end{document}